\documentclass[aps,preprint]{revtex4-2}

\usepackage{graphicx}
\usepackage{float}
\usepackage{amsmath}
\usepackage{amsfonts}
\usepackage{amssymb}

\begin{document}

\title{Geometric investigation of chaos unfolding in Hamiltonian systems}

\author{L. Salasnich$^{1,2,3}$  and F. Sattin$^{4}$}
\email{luca.salasnich@unipd.it; fabio.sattin@igi.cnr.it}
\affiliation{
$^{1}$Dipartimento di Fisica e Astronomia ``Galileo Galilei'' 
and CNISM, Universit\`a di Padova, Via Marzolo 8, 35131 Padova, Italy
\\
$^{2}$Istituto Nazionale di Ottica (INO) del Consiglio 
Nazionale delle Ricerche (CNR),
via Nello Carrara 1, 50125 Sesto Fiorentino, Italy
\\
$^{3}$Istituto Nazionale di Fisica Nucleare (INFN), Sezione di Padova, 
Via Marzolo 8, 35131 Padova, Italy
\\
 $^{4}$Consorzio RFX (CNR, ENEA, INFN, Universit\`a di Padova, 
Acciaierie Venete SpA), Corso Stati Uniti 4, 35127 Padova, Italy}

\begin{abstract}
In this work we revisit the geometric approach to chaos in Hamiltonian dynamics, by means of the Jacobi-Levi-Civita equation (JLCE).
We inspect numerically two low-dimensional dynamical systems; show that, along chaotic orbits, the exponential divergence between nearby trajectories quantified by the JLCE does not unfold in a continuous manner, rather is closer to a multiplicative discrete process: in correspondence of each turning point, where the trajectory bounces away from the boundary of the energetically allowed region, the relative separation increases sharply and abruptly.  We highlight through analytical and numerical arguments that the chaotic rather than regular nature of the trajectory is determined by the details of the scattering with the boundary, and interpret these results in terms of parametric resonance theory, and specifically the Mathieu equation.      

\hfill

{\it Keywords:} Hamiltonian systems; chaos; 
integrability; Riemannian manifolds; Jacobi-Levi-Civita equation; parametric resonance; Floquet theory;  Mathieu Equation

\end{abstract}

\maketitle

\section{Introduction}
The identification and characterization of ordered or disordered (chaotic) motion in dynamical systems is an old but still important topic. Knowing whether a dynamical system admits ordered or chaotic orbits and predicting the basins of stability may have deep implications for several  problems in physics and engineering. A few examples are: the assessment of stable conditions in exoplanets--a necessary requisite for the development of life; the stability of particle  beams in accelerators; the adiabatic {\it versus} non-adiabatic dynamics of fast charged particles in magnetized plasmas, the consequences upon their confinement being of paramount importance in nuclear fusion research. (As a matter of fact, one of the triggers of this work was the desire of identifying the basins of stability of single-particle motion near local magnetic minima, continuation of an enduring investigation about the boundaries of validity of gyrokinetics: the theoretical framework within which plasma dynamics in strong magnetic field is interpreted nowadays \cite{gyro23,cambon14,pfefferle15,sr19,nf21,sym21,pre23,sym23,chen24}). \\     
The natural environment of a Hamiltonian system is the phase-space one, where positions and momenta are treated on an equal footing, but a geometric approach to Hamiltonian dynamics exists and is known since long, as well; the Cartesian phase-space structure is there mapped into a curved Riemaniann manifold, where only spatial degrees of freedom remain, and Newtonian trajectories become the geodesics on this manifold (see \cite{aizawa72,churchill75,pettini93,pettini95,sola96,sola98,szczesny99,casetti00,saa04,awrejcewicz06,pettini07,horwitz07,calderon13}). Within this framework, the information about order and chaos is formally provided by the Jacobi-Levi-Civita Equation (JLCE), which quantifies the rate of divergence between two nearby geodesics. A convenient aspect of the JLCE is that it may naturally be formulated in a parallel transported frame, and therefore projected along directions parallel and perpendicular to the motion, only the latter ones being relevant for assessing the existence of chaos. In order to appreciate the importance of this feature, we compare it with the so-called Toda-Brunner-Duff (TBD) criterion \cite{toda74,brunner76} (although it was actually predated by Okubo \cite{okubo70}), which may be regarded as the phase-space homologous of the JLCE. The TBD approach involves the linearization of the Hamilton's  equations of motion for the phase-space {\it difference vector} ${\bf d}$ between two nearby trajectories, which we may write as ${\dot{\bf d}} = {\bf H}\cdot{\bf d}$. For a standard hamiltonian with separated kinetic and potential terms, $H = {\bf p}^2/(2 m) + V({\bf q})$,  where ${\bf q, p} $ are respectively the position and momentum degrees of freedom, the matrix {\bf H} is  
\begin{equation}
{\bf H} = \begin{pmatrix}  
                {\bf 0} & {\bf I} \\
                \frac{-1}{m} \frac{\partial^2 V}{\partial q_i \partial q_j}  & 0
                \end{pmatrix}
\end{equation}     
The eigenvalues of ${\bf H}$ provide the rate of divergence between nearby trajectories: if positive, they predict local exponential instability. The TBD criterion is, thus, extremely appealing: if working, it would be an analytic and local method for predicting the existence of unstable phase-space regions, and as such has been repeatedly employed \cite{enz75,nunez90,oloumi99,watabe95, kuvnishov02,kuvnishov08,li95,manfredi97, letelier11,ghosh14,chang14,beron19,shivamoggi22}. Unfortunately, it was also soon realized that it is an unreliable tool, since predicts ``false positives'', i.e., regions of hyperbolic instability in systems that can be rigorously proved to be regular, including one-dimensional systems like the pendulum, or two-dimensional systems like the anti-Hènon-Heiles system \cite{benettin77,pattanayak97,smirnov98}. The root of the problem is traced back precisely to the fact that the TBD criterion treats all directions on the same footing, and does not discriminate between relevant and irrelevant degrees of freedom \cite{sattin24}. 

It is useful at this point, before proceeding further, to recall the basics of the geometrization procedure. 
In the following we will restrict to two-dimensional autonomous Hamiltonian systems (i.e., whose phase-space is four-dimensional). While being the simplest non-trivial systems, as far as disordered dynamics is concerned, they are useful in practical cases, too: for instance, the dynamics of a single particle in prescribed fields is three-dimensional, and is reduced to two-dimensional in presence of one spatial symmetry, which often happens in strongly magnetized systems. \\
Let the system' hamiltonian be like above: $H = {\bf p}^2/2 + V({\bf q})$. For fixed energy $H = E$, the Jacobi metric tensor is        
\begin{equation}
g_{ij} =  (E - V({\bf q})) \, \delta_{ij} \equiv W \, \delta_{ij} \, , \quad i,j = 1,..., 2
\end{equation}
where $\delta$ is the Kronecker delta. The proper time $s$ is related to the metric tensor and to the physical time $t$ by
\begin{equation}
ds^2 = g_{ij}({\bf q}) \, dq^i dq^j = 2 W^2 \, dt^2
\end{equation}
Geodesics equations are
\begin{equation}
\frac{d^2 q^i}{ds^2}+ \Gamma^i_{jk} \, \frac{dq^j}{ds} \frac{dq^k}{ds} = 0
\label{eq:geodesics}
\end{equation} 
The coefficients $\Gamma^i_{jk}$ are the Christoffel symbols. Eqns. (\ref{eq:geodesics}), when converted to the original Cartesian space and physical time, are precisely the Newton equations, but motion along geodesics is free, meaning that its velocity is constant: 
\begin{equation}
g_{ij}  \frac{dq^i}{ds} \frac{dq^j}{ds} =   1
\end{equation}

The JLCE writes
\begin{equation}
\frac{\nabla^2 J^i}{ds^2} + R^i_{jkl} \frac{dq^j}{ds} J^k \frac{dq^l}{ds} = 0
\label{eq:jlce}
\end{equation}
where ${\bf J}$ is the separation vector between two nearby geodesics, $R^i_{jkl}$ the Riemann curvature tensor, ${\bf v} = d{\bf q}/ds$ the common flow. Finally, $\nabla /ds$ is the covariant derivative.  By projecting Eq. (\ref{eq:jlce}) onto a parallel-transported basis of vectors, it is possible to write it as a set of coupled differential equations,  linear in ${\bf J}$. In the two-dimensional case, the basis vectors are $ {\bf e}_\parallel = (dq^1/ds, dq^2/ds), \, {\bf e}_\perp = (-dq^2/ds, dq^1/ds)$, ${\bf J} = J_\parallel {\bf e}_\parallel + J_\perp {\bf e}_\perp$, and Eq (\ref{eq:jlce}) rewrites as
\begin{eqnarray}
&\frac{d^2 J_\parallel}{ds^2}& = 0  \label{eq:j2da}\\
&\frac{d^2 J_\perp}{ds^2}& +  {\cal R} \,  J_\perp = 0
\label{eq:j2db}
\end{eqnarray}   
The first of these equations restates that no acceleration (and, therefore, no instability) along the flow occurs. The second is the physically important one, since the physical mechanisms that produce regular or unstable dynamics are all embedded into the single Gauss scalar curvature parameter ${\cal R} = R_{1212}$, where $R_{ijkl} = g_{im} R^m_{jkl}$. \\
In particular, ${\cal R}$ is expressible in terms of the potential function as
\begin{equation}
{\cal R} = \frac{|\nabla V|^2}{W^3}+ \frac{\Delta  V}{W^2}
\label{eq:rgauss}
\end{equation}  
where $\nabla, \,\Delta$ are the Euclidean gradient and Laplacian operators respectively. \\
When ${\cal R} < 0$ over the whole, or at least a finite fraction of the, available space, then exponential separation of the trajectories arises. This is the hyperbolic instability scenario conceived originally by Anosov and Krylov. Quite often, though, ${\cal R}$ turns out to be posiitive everywhere. By inspecting the structure of  Eq.(\ref{eq:rgauss}) it is easy to be convinced that a necessary condition for ${\cal R} < 0$ is $\Delta V < 0$, i.e., the potential must feature local maxima or ridges. Since ${\cal R} > 0$ everywhere for notoriously chaotic systems, obviously chaos must be produced by some other mechanism, which was identified in the fluctuations of ${\cal R}$, in a way resembling the parametric resonance mechanism (\cite{param12}, chap. 10 in \cite{rajasekar16}). This corresponds to the situation where a system endowed with a natural frequency of oscillation, $\omega_s$, is forced by an external perturbation with a resonant frequency $\omega_f$: the paradigmatic case being provided by the Mathieu equation, where unstable motion can be triggered by arbitrarily small external perturbations, provided that $\omega_f$ resonates with $\omega_s$ according to the relation $\omega_s = \omega_f \times n/2$, $n$ being a positive integer. 

The ${\cal R}$-modulation mechanism must thus be at the basis of the generation of chaos in several dynamical systems. This is a success of the geometric approach but has represented, paradoxically, even a limit.  Unlike standard physics or engineering scenarios, where the forcing term is given independently of the dynamics,  here the curvature ${\cal R}$ is function of the system' coordinates, and therefore of the trajectory: in order to solve the JLCE, one has to feed it with a previously computed trajectory; thus, ultimately, must have beforehand some information about the kind of dynamics--which, obviously, makes useless the JLCE as a  tool for the chaos detection. The JLCE, in summary, has acquired a fundamental heuristic value in interpreting the nature of the system’ behavior, but at the price of losing any useful predictive power. The only loophole, so far, was to bypass the precise knowledge of ${\cal R}$ employing some approximate estimate based upon statistical arguments \cite{sola96,kandrup97,pettini07}. The recipe works only in very-high-dimensional systems, though.   \\  
In this paper we do not revolutionize the picture above, but mitigate it.  In the next section, we perform some numerical exercises employing a few classical Hamiltonian systems. We compute ${\cal R}(s)$ along both regular and chaotic trajectories, solving Eq. (\ref{eq:geodesics}). We find that, regardless the nature of the motion, the typical timetraces ${\cal R}(s)$ share several common traits: all are made by superposing narrow quasiperiodic peaks to a flat background. The peaked structures ${\cal R} = {\cal R}_{max}$ correspond to the reflections of the trajectories near the boundary of the energetically allowed region , where $ E - V \approx 0$ (from now on, ``the energy boundary'' for short), whereas the flat background is due to the almost-free-motion far from the boundaries, where $E \gg V$ and ${\cal R} \approx  {\cal R}_{min} = 2/E^2$. Since a constant ${\cal R}$ prevents exponential separation of trajectories in Eq. (\ref{eq:j2db}),  the difference between ordered and chaotic motion must be traced back to the details of the scattering at the energy boundary. Thus, we refine somewhat the widespread statement according to which chaos is a global property of a dynamical system. It is, but different phase-space regions cooperate with different roles in producing chaos. In section III, on the basis of an highly simplified analytic model of the ${\cal R}$ dynamics, we are able to build a stability diagram in the plane $({\cal R}_{min}, {\cal R}_{max})$ that discriminates quantitatively between ordered and chaotic trajectories. In summary, the novelty of the paper lies in pointing out the relevance of the dynamics near the energy boundaries with repect to the order/chaos nature of the system. \\
Finally, an appendix is devoted to a spin-off of the main theme: we show that, employing our results, it is possible to draw some inferences about the seemingly unrelated problem of nonadiabatic energization of charged particles during magnetic compression.  
     
\section{Numerical Experiments}
\begin{figure}
\includegraphics[width=9cm]{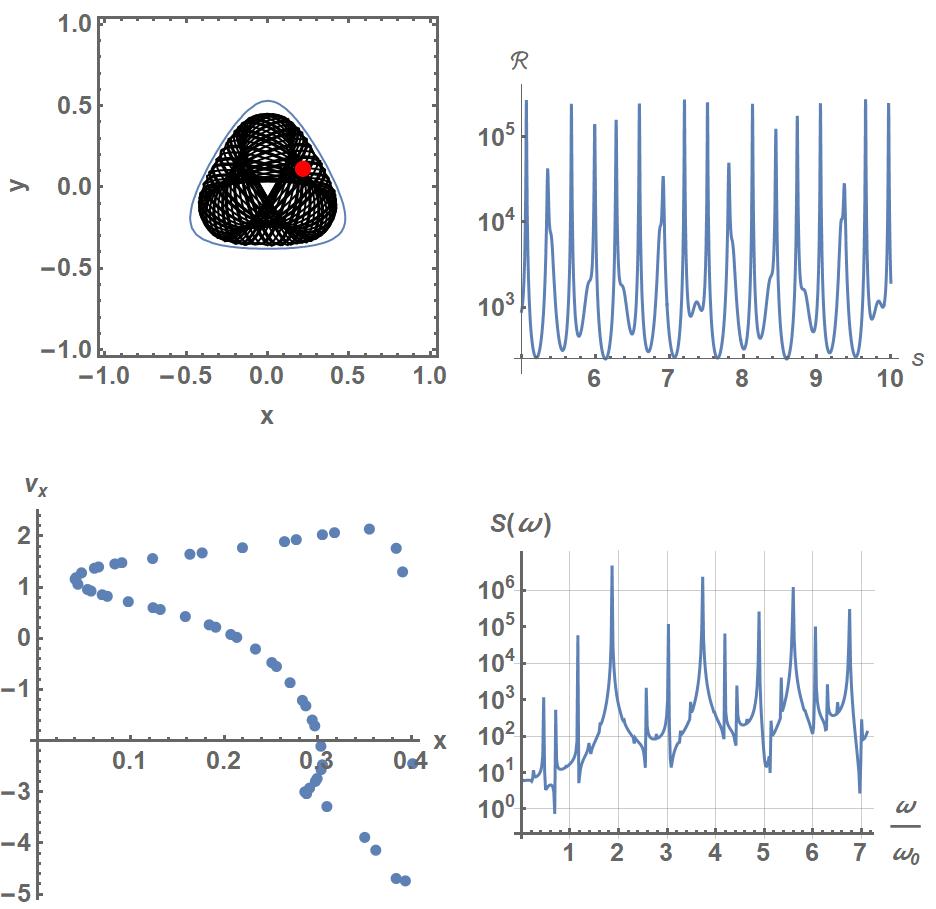}
\caption{An example of regular trajectory in the HH system. Left-top plot: the black curve is the trajectory in the $(x,y)$ plane computed using Eq. (\ref{eq:geodesics}) in the interval from $s= 0$ to $s = 30$. The blue curve is the energy boundary, $E = V= 1/11$. Initial conditions are: $x(0) = 0.2172, y(0) = 0.11205$ (shown as the red dot), $(dx/ds)(0) = -1.63898, (dy/ds)(0) = 2.49126$. 
Right-top plot  shows a part of the time trace ${\cal R}(s)$. Notice that ${\cal R}_{min} = 2/E^2 = 242$.
Left-bottom plot is the Poincaré plot, obtained taking the intersections of the trajectory with the line $y = 0, v_y > 0$. 
 The right-bottom plot shows the power spectra of $\sqrt{{\cal R}/2}$. Frequency is normalized to $\omega_0 = \sqrt{{\cal R}_{min}/2}= E^{-1}=11$.  }
\label{fig:1}
\end{figure}

\begin{figure}
\includegraphics[width=9cm]{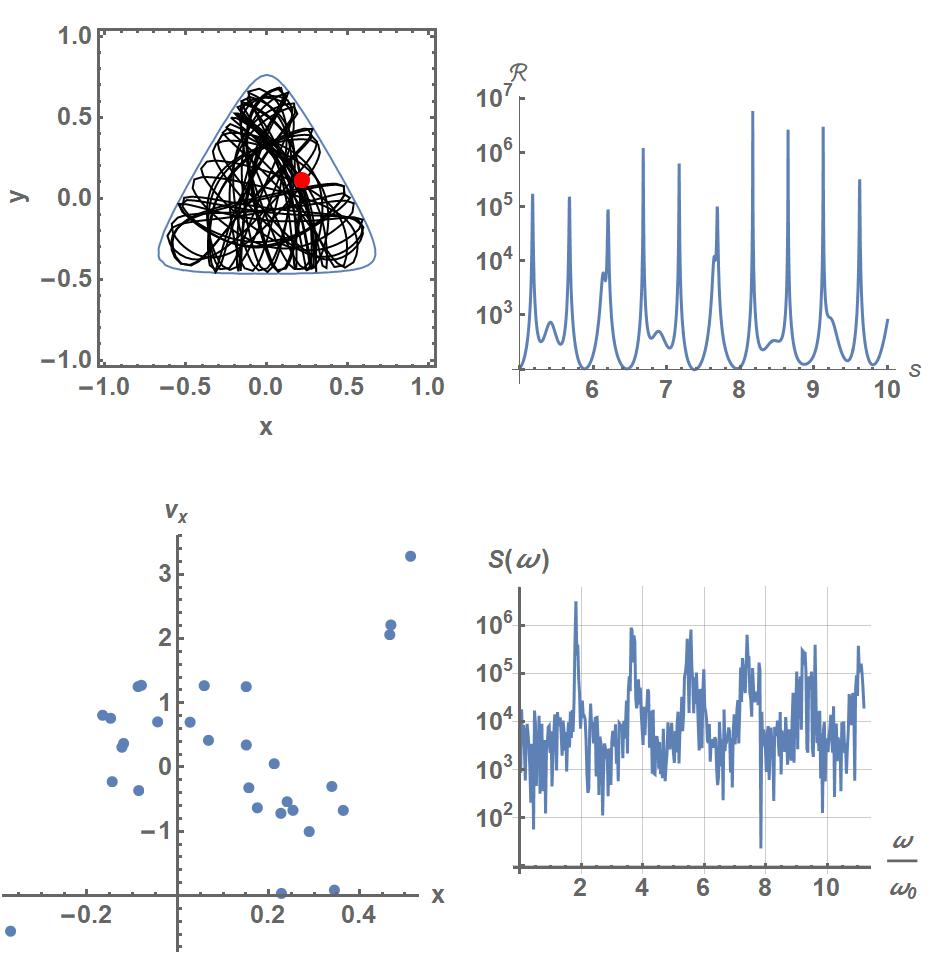}
\caption{A chaotic trajectory in the HH system. Initial conditions are the same of the previous figure, but now the energy is $E = 1/7$, ${\cal R}_{min} =98$ . }
\label{fig:2}
\end{figure}

\begin{figure}
\includegraphics[width=9cm]{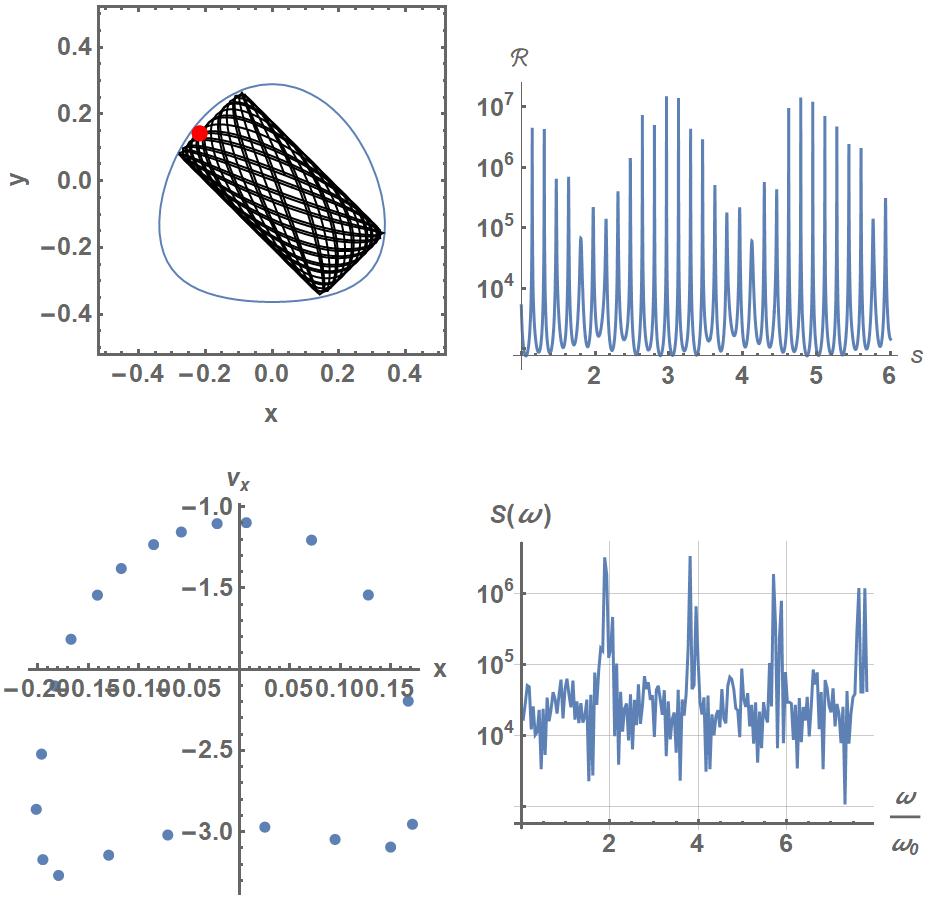}
\caption{The content of the figure is the same of Figs. (\ref{fig:1}, \ref{fig:2}), relative to a trajectory of the aHH system.  Here, $E = 1/20$, ${\cal R}_{min} =800$,  and initial conditions are: $x(0) = -0.2172, y(0) = 0.14205, (dx/ds)(0) = -2.90459, (dy/ds)(0) = -7.01845 $. }
\label{fig:3}
\end{figure}

We will consider two reference potentials. The first one is the Hénon-Heiles (HH) potential: a workhorse of any study of chaos in dynamical systems:
\begin{equation}
V  = {1 \over 2} \left( x^2+ y^2 \right) + x^2 y - { y^3 \over 3}
\label{eq:hh}
\end{equation}
Besides  this, we consider also the anti-Hénon-Heiles (aHH) potential:
\begin{equation}
V  = {1 \over 2} \left( x^2+ y^2 \right) + x^2 y + { y^3 \over 3}
\label{eq:ahh}
\end{equation}
Despite the apparently tiny difference consisting of just the sign in front of the last term, the two systems behave in fairly different ways: the HH is a regular system at low energy, $ E \leq 1/12$; then a fraction of the trajectories becomes chaotic by increasing $E$, until $E = 1/6$, where the fraction reaches the unity. Beyond this threshold, the trajectories are no longer bounded. The aHH system, conversely, is regular at any energy \cite{aizawa72,codaccioni82,pattanayak97}. Indeed, the two systems may be thought of as particular cases out of one single parametric class of hamiltonians. The appearance and disappearance of integrals of motion in correspondence of the variation of these parameters is a field of study \cite{smirnov98,contopolous25,tzemos25}.  \\
In  figs. (\ref{fig:1},\ref{fig:2},\ref{fig:3}), we provide a few typical samples of numerically integrated trajectories. For each example, we plot: the trajectory; the Poincaré plot, produced taking the intersections of the trajectory with the $y = 0$ line; the Gauss scalar curvature ${\cal R}$ computed along the trajectory, and the frequency spectrum of the orbit. Notice that the trajectories were computed by integrating the geodesics equation (\ref{eq:geodesics}). For a limited number of cases, results were compared and validated against the solution of Hamilton equations obtained using a symplectic integrator \cite{atela92}.

\begin{figure}
\includegraphics[width=9cm]{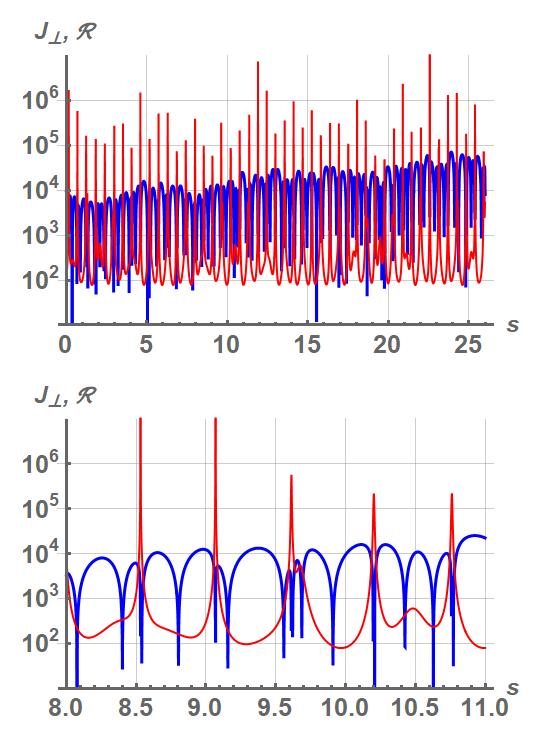}
\caption{Upper plot: the red curve is ${\cal R}$ from a chaotic trajectory of the Hénon-Heiles system; the blue curve is the deviation $J_\perp$ (arbitrarily rescaled in order to fit the window) computed from Eq. (\ref{eq:j2db}), feeded with  ${\cal R}$. The lower plot shows an expanded view of the same curves, over a restricted time span.  
The energy is $E = 1/6.25$. Initial conditions are: $x(0) = -0.0172, y(0) = 0.24205, (dx/ds)(0) = -2.90459, (dy/ds)(0) = 3.01845 $. }
\label{fig:chao}
\end{figure}

\begin{figure}
\includegraphics[width=9cm]{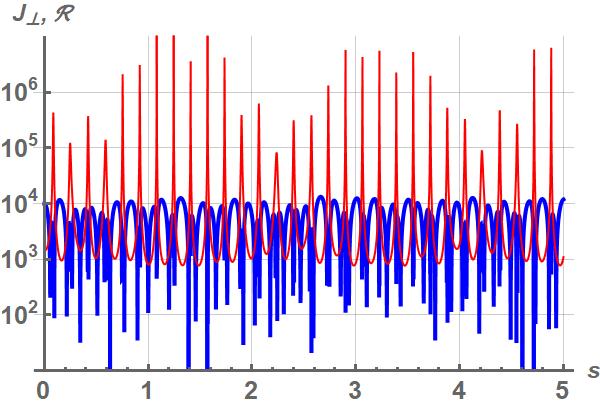}
\caption{The red curve is ${\cal R}$ from a regular trajectory of the anti-Hénon-Heiles system; the blue curve is the deviation $J_\perp$ (arbitrarily rescaled in order to fit the window) computed from Eq. (\ref{eq:j2db}), feeded with  ${\cal R}$. 
The energy is $E = 1/20$. Initial conditions are: $x(0) = 0.12, y(0) = 0.03, (dx/ds)(0) = -2, (dy/ds)(0) = 3 $.}
\label{fig:reg}
\end{figure}

The curve ${\cal R}(s)$ looks like as made by two or three superposed contributions: (i) a constant floor, corresponding to the ``quasi-free-motion'' part of the orbit, where ${\cal R}  \approx {\cal R}_{min} = 2/E^2$. (ii) On top of this background, narrow and tall peaks are superposed, corresponding to the reflection of the trajectory near the energy boundary: by looking at Eq. (\ref{eq:rgauss}) we notice that, ${\cal R} \propto (E-V)^{-3} \to \infty$ when $V \approx E$. These peaks appear at a fairly regular cadence, regardless the stability of the trajectory: The Fourier spectra are quite similar amongst all cases, with equispaced resonances. The quasi-periodicity of ${\cal R}$ even with chaotic trajectories is quite surprising: on first thought, one may conjecture that ${\cal R}$ would inherit the irregular behaviour of the trajectory \cite{kandrup97}.  (iii) A second kind of peaks may appear,  too: these latter are broader and smaller, but appear with the same frequency as the first kind. They correspond to grazing reflections, where only a small fraction of the velocity is aligned along the potential gradient. The presence of this second kind of peak appears characteristic of the Hénon-Heiles system alone, they are absent or hardly detectable in the aHH model. \\
Feeding into Eq.(\ref{eq:j2db}) the curvature ${\cal R}$ computed in the previous step, we may compute the deviation $J_\perp$. We present two instances of solution in Figs.  \ref{fig:chao}, \ref{fig:reg}, corresponding respectively to one Hénon-Heiles chaotic trajectory, and one anti-Hénon-Heiles regular one. As expected, we recover $J_\perp$ growing exponentially in the first case, while remains constant in the second case. However, we are talking of the average trends; The full behaviour of $J_\perp$ is oscillatory, with an instantaneous frequency $\sqrt{  {\cal R} / 2}$.  It is possible to appreciate from the bottom plot of Fig. \ref{fig:chao} that, over smaller time scales, the average growth manifests itself as a succession of small increments (which in some cases, can even be temporarily {\it decrements}) in correspondence of the intersection with the peaks of ${\cal R}$.

The modulation of ${\cal R}$ is expressed thus through these peaks, which are defined in terms of  their defining parameters (height, width, shape, periodicity, ...). The purpose of the next section is to guess how these parameters discriminate between ordered and  chaotic orbits.  

\section{An analytical solution for the JLCE}
\begin{figure}[H]
\includegraphics[width=9cm]{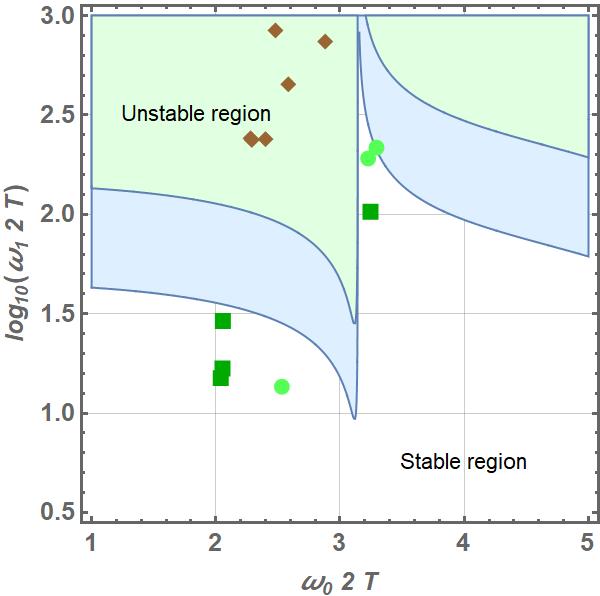}
\caption{Ince-Strutt stability diagram for the periodic-rectangle model of ${\cal R}$. Shaded areas identify the unstable regions, where $max(|\lambda|) >1$: the blue area corresponds to $d/T = 2\times 10^{-4}$,the green area to $d/T = 20\times 10^{-4}$.  Superposed, are the positions in  this plane of some sample trajectories. Respectively, ordered trajectories in the anti-Hénon-Heiles problem (light green dots); ordered trajectories in the Hénon-Heiles problem (dark green squares); chaotic orbits in the Hénon-Heiles problem (brown diamonds).  All numerically measured $d/T$ ratios for these trajectories fall  in the semi-interval $d/T \ge 2 \times 10^{-4}$. }
\label{fig:5}
\end{figure}

The scalar curvature $ {\cal R}(s)$ takes the form of a flat background with superposed very narrow tall peaks. As an analytical approximation, we replace ${\cal R}$ with a constant term, plus a periodic superposition of tall rectangles, like in Fig. \ref{fig:4} :
\begin{figure}[H]
\includegraphics[width=9cm]{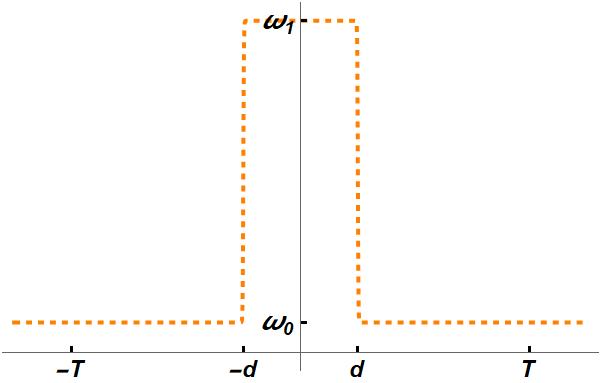}
\caption{Dashed orange curve: sketch of the shape for the curvature function. $\sqrt{{\cal R}/2}$ jumps from $ \omega_0$ to $ \omega_1$ within a time interval $-d  <s < d$. The pattern repeats periodically with period $2 T$.  }
\label{fig:4}
\end{figure}
We assume that the peaks are periodic, with period $2 T$; that is, they appear at $s = 0, 2 T, 4 T, ...$. With this choice for ${\cal R}$, the JLCE becomes the equation for a harmonic oscillator where the oscillation frequency jumps between two constant values, and its solutions are trivially trigonometric functions. 
The stability of the system is assessed using the traditional Floquet method. The JLCE is solved from $ s = -T$  as  an initial-value problem, where the initial values are provided by the two fundamental solutions:
\begin{equation} 
 y_A = \cos \left[\omega_0 \left( s +T \right) \right] \; , \; y_B = \omega_0^{-1}  \sin  \left[\omega_0 \left( s +T \right) \right]
\end{equation}
W integrate them up to $s = T$, and assemble there the matrix
\begin{equation}
M = 
\begin{pmatrix}
 y_A & y_B \\
       y_A'& y_B' 
 \end{pmatrix}
 \label{eq:m}
 \end{equation} 
(The prime stays for the derivative). The eigenvalues $\lambda$ of $M$ establish the character of the solution: if $|\lambda| > 1$ we have a multiplicative process where the outgoing solution (i.e., at $s = T$) has a larger amplitude than the ingoing one, at $s = -T$. If repeated periodically, this leads to an overall exponential growth: after $N$ peaks (turning points), the initial separation has increased by a factor $ |\lambda|^N = \exp ( N \ln |\lambda|) = \exp \left[ (2 T N) \, \ln |\lambda| (2 T)^{-1} \right] = \exp( s \, \lambda_L)$, an exponential trend where $s = 2 T N$ is the time, and $\lambda_L = \ln |\lambda| (2 T)^{-1}$ the Lyapunov exponent.  \\
The eigenvalues are computable analytically in terms of the three normalized parameters $ 2 \, T \, \omega_0, 2 \, T \, \omega_1 , d/T$.  Their explicit expression is cumbersome, so is not provided here, however it is easy to draw a stability diagram, where one of the three parameters is kept fixed and the other two allowed to vary.  In Fig. \ref{fig:5} we plot it for two values $d/T = 2\times 10^{-4}, d/T = 20\times 10^{-4}$:the colored areas (respectively blue and green) mark the unstable regions in the $(\omega_0, \omega_1)$ space. Some limiting cases can be worked out explicitly: It is possible to show that in correspondence of the vertical lines at $2 \, T \, \omega_0 = n \pi, n = 1,2,...$, $|\lambda| = 1 + {\cal O}(d^2)$; besides $|\lambda| \to 1$ when $ d \to  0$.  
In order to test the reasonableness of the simplified periodic-rectangle model, we compare its predictions against numerical results.
We have produced a few sample trajectories, for both the HH and aHH systems. For each trajectory, 
the period $2 T$ is estimated as the average distance in $s$ between two peaks, and $\omega_0 = E^{-1}$. Finally, $\omega_1$ is numerically evaluated from the time series of $\sqrt{{\cal R}/2}$ between $s = 0$ and $s = s_{max}$  as the average height of the peaks. The average width $2 d$ of the peaks, finally is estimated as $2 d = (\int_0^{s_{max}} (R/2- \omega_0^2) ds)/(N_p (\omega_1-\omega_0))$ , and $N_p$ is the number of the peaks.  
We add to the stability siagram of Fig. \ref{fig:5} the points $(\omega_0, \omega_1)$ computed according to the above algorithm, for a few trajectories from the regular anti-Hénon-Heiles system (light green  symbols), Hénon-Heiles system (regular trajectories--dark green symbols--and chaotic trajectories--brown symbols). In all cases, the regularity or less of the trajectory was estimated by inspection of its Poincaré plot. Encouragingly, all points locate in the region of the stability diagram that is consistent with the character of its motion.  \\
Qualitatively, the stability diagram (\ref{fig:5}) may be looked at affine to that produced by the Mathieu equation, where unstable regions appear as tongues of increasing width as some parameter ($\omega_1 $ in our case) is let to increase \cite{butikov18}.
 
\section{Conclusions}  
The main result out of this study is the acknowledgment of the role of the dynamics in the neighbourhood of the energy boundary.  First of all, we have argued that the impact of this dynamics can be packed effectively into a very coarse model, in terms of few parameters. Secondly, the stability diagram (\ref{fig:5}) produced using this model hints that any dynamical system might be made unstable if reflections are allowed to occur sufficiently close to the energy boundary. TWe may guess that stable dynamical systems are distinguished from chaotic onesby the fact that, in the former, orbits keep an appreciable component along iso-potential curves (equivalently, motion purely along the gradient of the potential is prevented). This is far from an explanation for the chaos/order duality, still represents potentially another starting point from which to tackle the problem. \\
We point out that the role of the energy boundaries upon the nature of the dynamics was already addressed by some authors. In particular, it was taken into consideration the impact of the convexity or concavity of the boundary on the stability of the motion \cite{churchill75,pattanayak97,saa04}. There are obvious affinities with billiard systems, where the presence of chaos is critically related to the details of the curvature of the energy boundary \cite{buninovich18}. However, ultimately the behaviour of smooth dynamical systems near the boundaries is much more complex, the production of chaotic dynamics cannot be associated to simple dispersing or defocussing mechanisms.  \\
Concerning the limitations as well as the open points left by our study: they are numerous, of course. We have addressed only few two-dimensional systems. We are quite confident that our conclusions hold generically for this class of systems, but to which extent they may be exported to higher-dimensional systems (e.g., to the three-dimensional systems, which are physically at least as much important as the two-dimensional ones) is an open question. On the other hand, the results obtained with two-dimensional systems might be useful for guiding our understanding about higher-dimensional ones: in systems of dimension larger than two the simple scalar JLCE becomes a coupled system of equations, since we have the equivalent of $J_\perp$ for each direction perpendicular to the flow. The analysis may result more complicated, but its interpretation might still be provided by the simple model produced in our section III.           

\section*{Acknowledgments} The authors wish to thank  D.F. Escande, M. Pettini, I. Predebon, L.A. Bunimovich for useful discussions, and the anonymous referees for precious comments and questions.

\section*{Author contributions}  Conceptualization: L.S. and F.S.;
Data curation: L.S. and F.S.;
Formal analysis: F.S.;
Investigation: L.S. and F.S.;
Methodology: L.S. and F.S.;  
Software: F.S;
Visualization: L.S. and F.S.;
Writing – original draft: L.S.and F.S.;
Writing – review and editing: L.S. and F.S.

\section*{Appendix}
We note that Eq. (\ref{eq:jlce}) is the equation of motion for a harmonic oscillator with time-varying frequency, or a charged particle in a magnetic field variable in time. In particular, the waveform of ${\cal R}$  encountered in figs. \ref{fig:1},\ref{fig:2},\ref{fig:3}, reproduces the dynamics of a fast magnetic compression. The energization of charged particles during magnetic field compression is a topic that received some interest, due to its relevance in natural scenarios (magnetic reconnections), but also laboratory plasmas \cite{krivets70,persson93,agren96,onofrio26}. 
The relevant dimensionless parameter defining the nature of the dynamics is $\sigma = |\omega'|/\omega^2$, where the prime stands for the time derivative and $\omega$ is the variable frequency. Expressed in terms of the magnetic field, $\omega = q B/m$, where $q,m$ are the charge and mass of the particle. When $\sigma \ll 1$, motion is adiabatic, conservation of the adiabatic invariant $\mu = K_\perp /(2 B) $--where $K_\perp$ is the kinetic energy due to the motion perpendicular to ${\bf B}$--entails that the kinetic energy $K$ varies with $B$ according to $K(t_2)/K(t_1) = B(t_2)/B(t_1)$ (the motion along the field being unaffected). Conversely, $\sigma \gg 1$ corresponds to the case where the variation of the magnetic field occurs over times scales much shorter that the Larmor period: the particle has not time to complete a single rotation before the field has substantially changed, and adiabaticity is broken down. It is possible to show (see \cite{krivets70}) that, in this case,  $K(t_2)/K(t_1) = (B(t_2)/B(t_1))^\nu$, with $2 >\nu > 1$, and $\nu \to 2$ when the field variation is infinitely sharp.  However, the breakdown of the adiabatic dynamics brings another consequence, if we allow for a cyclical modulation of $B$, i.e., at some $t_2 > t_1$ the field returns to its initial value: $B(t_2) = B(t_1)$. In the adiabatic case $K(t_2) = K(t_1)$, whereas in the nonadiabatic case $K(t_2) \neq K(t_1)$ (and, for most orbits, $K(t_2) > K(t_1)$). \\
We have computed numerically $\sigma$ for the orbits studied in the present work, and found that it takes always moderate values: $\sigma \leq 0.7$, both for regular and chaotic trajectories. This entails that another path to a substantial particle energization through magnetic field compression is possible: rather than through a single strongly non-adiabatic stroke, it may be achieved via several periodic mildly non-adiabatic ones. Although the gain after a single cycle is not much larger than unity, the overall growth is exponential with the number of cycles.

\end{document}